# Nowcasting Earthquakes with QuakeGPT: Methods and First Results

by

John B Rundle[1,2], Geoffrey C. Fox[3], Andrea Donnellan[4], Lisa Grant Ludwig[5]

[1] University of California, Davis, CA USA

[2] Santa Fe Institute, Santa Fe, NM, USA

[3] University of Virginia, Charlottesville, VA, USA

[4] Jet Propulsion Laboratory
California Institute of Technology, Pasadena, CA, USA

[5] University of California, Irvine, CA, USA

## Abstract

Earthquake nowcasting has been proposed as a means of tracking the change in large earthquake potential in a seismically active area. The method was developed using observable seismic data, in which probabilities of future large earthquakes can be computed using Receiver Operating Characteristic (ROC) methods. Furthermore, analysis of the Shannon information content of the earthquake catalogs has been used to show that there is information contained in the catalogs, and that it can vary in time (Rundle et al., 2023a, 2024c). Here we discuss a new method for earthquake nowcasting that uses an AI-enhanced deep learning model "QuakeGPT" that is based on an attention-based science transformer adapted for time series forecasting. Such dot product attention-based transformers were introduced by Vaswani et al. (2017), and are the basis for the new large language models such as ChatGPT. To use these science transformers, they must first be trained on a large corpus of data. A problem is that the existing history of reliable earthquake catalog data extends back in time only a few decades, which is almost certainly too short to train a model for reliable nowcasting/forecasting. As a result, we turn to earthquake simulations to train the transformer model. Specifically we discuss a simple stochastic earthquake simulation model "ERAS" (Earthquake Rescaled Aftershock Seismicity) that has recently been introduced (Rundle et al, 2024c). The ERAS model is similar to the more common "ETAS" models, the difference being that the ERAS model has only 2 novel, adjustable parameters, rather than the 6-8 adjustable parameters that characterize most ETAS models. Using this ERAS model, we then define a transformer model and train it using a long catalog of ERAS simulations, then apply it to an ERAS validation dataset with the transformer model. In this paper, we describe this new method and assess the applicability to observed earthquake catalogs for use in nowcasting/forecasting.

## Plain Language Summary

Earthquake nowcasting was proposed as a means of tracking the change in the potential for large earthquakes in a seismically active area, using the record of small earthquakes. The method was developed using observed seismic data, in which probabilities of future large earthquakes can be computed using machine learning methods that were originally developed with the advent of radar in the 1940s. These methods are now being used in the development of machine learning and artificial intelligence models in a variety of applications. Additionally, in recent times methods to simulate earthquakes using the observed statistical laws of earthquake seismicity have been developed . Here we describe a new approach to earthquake nowcasting using a model "QuakeGPT" that combines earthquake simulations with science transformers, a type of deep learning model using the concept of dot product attention that was introduced by Vaswani et al. (2017). In this paper we describe this new approach and assess its advantages and challenges, and discuss future directions for these methods. We note that the approach is general, and may also be applied to many other problems in time series forecasting.

## Key Points

- Earthquake nowcasting tracks the change in the potential for large earthquakes, using information contained in seismic catalogs
- We develop a new method for nowcasting by using stochastic earthquake simulations to train an AI-enhanced deep-learning neural network model
- We find that the method has considerable promise, but needs substantial additional development

## Introduction

Earthquake nowcasting (Rundle et al., 2016; 2019; 2021a,b,c; 2022;a 2023; Rundle and Donnellan, 2020; Pasari 2019; 2020; 2022; Pasari and Mehta 2018; Chouliaras et al., 2023) is a relatively new method that uses elements of machine learning to track the current state of the potential for large earthquakes, as well as the recent past and near future. Our previous research in earthquake nowcasting has been based on the Receiver Operating Characteristic (ROC) method that was developed with the invention of radar in 1941 relating to the observation of "signals" associated with reflections, or "events" (https://en.wikipedia.org/wiki/Receiver_operating_characteristic).

The fact that one can compute a probability for large earthquakes (Rundle et al. , 2023) implies that there is information (Shannon, 1948) contained within seismic catalogs. While random clustering can occur naturally as a result of Poisson statistics (Gross and Rundle, 1998), it can be shown that the information arises from the "non-Poisson", scale-invariant power-law clustering of aftershocks (Rundle et al., 2024c). This non-Poisson clustering typically arises following large earthquakes, or as swarms or "bursts". More specifically, it was found that information is associated with the relative quiescence that follows aftershock activity, meaning that removing aftershocks obscures the boundary between an "active" phase vs. a "quiescent" phase.

The purpose of this paper is to describe alternative methods for earthquake nowcasting that make use of science transformers, which are the basis for Large Language Models (LLMs) such as ChatGPT (e.g., Rundle, 2023; Chang et al., 2023; Yang et al., 2023; Fox et al., 2022). In these science transformers, models are trained on a large catalog of data, allowing the transformer to build a predictive model based on the concept of query, key and vector inputs to an encoder-decoder architecture combined with the concept of dot-product attention (i.e., Vaswani et al., 2017). LLMs are trained on large collections of facts, internet sites, documents, computer codes, images, and other data. We adapt this science transformer architecture to analyze time series of earthquakes and more general time series for predictive purposes.

In the earthquake application, we do not generally have access to the volume of earthquake catalog data that we require, since these catalogs only extend back a few decades in time with reliable data. As a result, we must turn to other means of training the transformer models, specifically, by the use of extensive simulations of seismic activity. For this reason, we are initially using stochastic earthquake simulations, which are physics-informed (but not physics-based) models that are constructed using the observational laws of earthquake seismicity.

The basic idea in this paper is to show how we can train the transformer model on a long sequence of earthquake data in a simulation catalog, and then apply the trained transformer model to a validation data set to evaluate the accuracy of the nowcast. In this paper, the validation data set is also a simulated catalog with the same simulation parameters. We then predict values of the nowcast time series beyond the end of the validation time series, which is what will be eventually done using observed data. We defer application of the transformer model to the observed California data to a later publication.

To summarize our results: This paper is organized as follows. We begin with a brief description of our current nowcasting methods. We then discuss the stochastic simulation model "ERAS" (Earthquake Rescaled Aftershock Simulation) that we use to train the transformer. We then summarize the implications for Shannon information entropy that we have published elsewhere. Next we discuss the construction of an initial dot product attention-based science transformer "QuakeGPT" to be used in nowcasting. We also describe the queries, keys and values that are used in the transformer, and then discuss alternative approaches under development. We next train a model using several thousand years of simulation data, and show how the model can fit an independent validation time series, predicting one-step walk-ahead values for the time series. Using the predicted values as input, we show that it is possible to predict future values of the time series beyond the end of the validation set, albeit with increasing levels of prediction error. Finally we assess the overall prospects and future promise for use of QuakeGPT as a means of anticipating the occurrence of future large earthquakes.

## Continuous Time Earthquake Nowcasting with ROC Methods

**Method.** In previous papers, we have described the nowcasting model, which tracks the current state of the complex dynamical earthquake system in space and time (e.g., Rundle et al, 2016; 2022a). Briefly, the method uses data in the seismically active region around Los Angeles (Figure 1). We proceeded by computing the current large earthquake potential state (EPS), which is defined as the exponential moving average (EMA) of the inverse monthly rate of small earthquakes. We then defined a "signal", which is the current value of the EPS above or below a selected threshold, and an "event", which is the occurrence or non-occurrence of a large earthquake within a selected future time window $T_W$. In (Rundle et al., 2022a; 2023), the value of $T_W$ is typically 1-3 years.

To implement the ROC method, we establish an arbitrary threshold value for the EPS curve, and categorize the "signals" and "events" into 4 categories. These are:

- If the "signal "is above the threshold, and a large earthquake *does occur* within the following $T_W$ years ("event"), the "signal" is a True Positive (TP).
- If the "signal "is above the threshold, and a large earthquake *does not occur* within the following $T_W$ years ("event"), the "signal" is a False Positive (FP).
- If the "signal "is below the threshold, and a large earthquake *does occur* within the following $T_W$ years ("event"), the "signal" is a False Negative (FN).
- If the "signal "is below the threshold, and a large earthquake *does not occur* within the following $T_W$ years ("event"), the "signal" is a True Negative (TN).

The initial threshold value is established at the lowest value of the EPS curve, all points on the timeseries curve are evaluated, and the confusion matrix is built for that value of the threshold. The threshold is then incrementally increased, and a new confusion matrix is built for that new threshold. This process continues until the maximum value of the EPS curve is reached. Typically one chooses several hundred threshold values, each resulting in its characteristic confusion matrix. From these confusion matrices, the ROC curve is then constructed by plotting the True Positive Rate (TPR), against the False Positive Rate (FPR), where:

$$TPR = \frac{TP}{TP + FN}, \; FPR = \frac{FP}{FP + TN} \quad (1)$$

It is important to note that all quantities, TP, FP, TN, FN, are functions of the threshold values at which they are computed, thus there are as many confusion matrices as there are threshold values. All confusion matrices have the same total number of entries, which is the number of points on the nowcast curve.

We can also compute the Positive Predictive Value, or Precision (PPV), which is defined as:

$$PPV = \frac{TP}{TP + FP} \quad (2)$$

The precision can be regarded as the probability of a future large earthquake occurring within the following $T_W$ years if the EPS curve is at, or above, a reference value on the EPS curve.

**Exponential Moving Average (EMA).** We considered a time series $T(t_j)$ representing the number of small earthquakes occurring in a month, that is indexed at regular monthly intervals $j$ = 1,...,$J$. As described in Rundle and Donnellan (2020), the EMA averages over the preceding times with an exponentially diminishing weight for more remote times in the past. More specifically:

$$T_{EMA}(t) = \begin{cases} T(t_0) & t_j = 0 \\ \alpha T(t_j) + (1-\alpha)T(t_j - 1) & t_j > 0 \end{cases} \quad (3)$$

Here, $\alpha \leq 1$ represents the amount of weighting decrease. A higher value of $\alpha$ discounts older observations faster. The most common choice for $\alpha$ is in terms of a number N, and is computed as $\alpha$ = 2/(1+N). The value of N, together with a second parameter, $R_{min}$, is chosen using the Receiver Operating Characteristic skill score, which is used as the loss function in a supervised learning scheme.

These quantities are defined using the state variable time series:

$$\Theta(t) \equiv \text{Log}_{10}(1 + \text{Monthly Number}) \quad (4)$$

Here we adopt the future time window $T_W$ = 3 years and classify all points on $\Theta(t)$ for a given threshold value into the 4 possibilities, TP, FP, FN, TN using the confusion matrix. The threshold is then varied over many values to produce an ensemble of confusion matrices. This then yields $TPR(T_h)$, which is a function of the threshold value $T_h$, as an implicit function of $FPR(T_h)$ as described above.

**Nowcasting with California Earthquakes.** We applied the EMA to the monthly rate of small earthquakes in the region, using $N_{EMA}$ = 36 months, in this case. Then, a correction to optimize the ROC skill of the nowcast was applied, partially to account as well for observational catalog incompleteness. In that Rundle (2022a) we used earthquakes having $M \geq 3.3$ since 1960.

After applying the EMA and small earthquake correction, the result is a smoothed version of the monthly rate of small earthquakes, which resembles an "upside-down" version of the earthquake cycle of stress accumulation and release, as has been pointed out in Rundle et al. (2022a). Inverting this smoothed seismicity curve, we obtain the nowcast timeseries curve, examples of which are shown in Figures 1 and 2, in which seismic activation is at the bottom of the diagram (many earthquakes such as aftershocks), and quiescence is at the top. It can clearly be seen that quiescence is the precursor to large earthquakes (e.g., Rundle et al., 2011).

The resulting temporal ROC is shown in Figure 1 for the time interval 1970-2022. The red curve is the ROC curve for $\Theta(t)$. The diagonal line from lower left to upper right is the "no skill" line. We also show as cyan lines ROC curves for 200 random time series by sampling randomly from $\Theta(t)$ with replacement (bootstrap method). The dashed black lines represent the standard deviation of the random time series. The entire process is shown schematically in Figure 1.

The final result of this process is shown in Figure 2, which is shown for a time just prior to the 2010 M7.2 El Mayor Cucupah earthquake. Note that, along with the time series, we also can plot the contours of likely locations for future large earthquakes. Figure 2 is a frame from a movie that is available online at (Rundle Movie, 2024a) or via Rundle et al., (2022) under the heading "Open Research".

## Stochastic Simulation: Earthquake Rescaled Aftershock Seismicity "ERAS"

**Stochastic Simulations of Seismicity.** As an initial step in the feasibility study of a QuakeGPT model, we build stochastic earthquake simulations to train our machine learning model. The most common example of this type of model is the Epidemic Type Aftershock Sequences (ETAS) formalism. These ETAS models are based on using the observed statistical laws of earthquake seismicity to build the rate for non-homogeneous Poisson models that can be used to generate space-time seismicity catalogs (Helmstetter and Sornette, 2003; Zhuang, 2011; Zhuang et al., 2012; Zhuang et al., 2015; Mancini et al., 2021; Veen et al., 2008; Lombardi, 2015; Mancini and Marzocchi, 2023; Hardebeck, 2013; Seif et al., 2017; Varotsos et al., 2011, 2014, 202).

The observational laws of interest are the Gutenberg-Richter (GR) magnitude-frequency relation; the Omori-Utsu (OU) aftershock time decay law; an analogous Omori-Utsu law for spatial aftershock migration away from the epicenter; and the Bath's law that governs the magnitude of the largest aftershock relative to the mainshock magnitude.

By themselves, these observed laws involve the scaling exponents $b$ (GR); $p$ and a parameter $c_p$ (OU-time); and $q$ together with a corresponding parameter $c_q$ (OU-space). In addition, there are additional parameters and relations that are assumed, for example an "earthquake productivity" relation having 2 parameters ($K$ and $\alpha$), and a background seismicity rate $\eta(x,y)$.

The typical process of fitting all these parameters to a catalog is usually carried out by maximizing a log-likelihood value involving the difference between the parameterized rate, and the rates determined from the catalog. Because all these parameters are involved in the overall log-likelihood equation, they are necessarily correlated, so that changing the value of one parameter necessarily involves changes in the values of other parameters. This makes the fitting process time-consuming and subject to correlation errors.

**ERAS: General Approach.** In this paper, we develop a different and simpler stochastic simulation model, an Earthquake Rescaled Aftershock Seismicity ("ERAS") model, which involves only 2 free, independent, and uncorrelated parameters, one for time, one for space. The first of these parameters, $f$, determines the Omori-Utsu aftershock time decay relation with exponent $p$. The second parameter, $g$, governs the Omori-Utsu aftershock spatial migration scaling with scaling exponent $q$.

These two parameters are determined by by fitting the ($p$, $q$) values observed from stacked aftershock seismicity using a simple ordinary least squares approach, which, as is well known from the Gauss-Markov theorem, is equivalent to a maximum likelihood approach. The model is constructed using the standard statistical relations of magnitude-frequency scaling, aftershock scaling in time and space, and Bath's law. This process is similar to the ETAS models, but without the additional 4-5 free parameters that are used in the ETAS models. We then applied the nowcast method to these ERAS models and discussed the results regarding information origin (Rundle et al., 2024c). Here we give a brief summary for constructing an ERAS catalog. A fuller description of the details of catalog construction is given in the appendix to this paper.

We build the simulated catalog in 3 steps. We first generate simulated catalogs having random (Poisson) interevent earthquake times to find a candidate catalog that has a *b* value within the observed margin of fitting error. Magnitudes of these events are drawn from a Gutenberg-Richter distribution. Next, since a (fractal) power law implies a geometric scaling or recursion, e.g. similar to the von Koch curve and other fractal curves (e.g., Turcotte, 1997), we apply a geometric clustering algorithm in time using the single parameter *f*. We then search for a value of *f* that produces a match to the observed *p* value within the margin of fitting error for an ensemble of stacked mainshocks.

For the spatial (fractal) power law scaling, we assume that aftershocks migrate away from the mainshock epicenter by a random walk, which is known to have scaling (power law) distribution properties. We then analyze the observational catalog to determine the ratio of random walk step sizes in latitude/longitude, and introduce a multiplicative scaling parameter *g*. Finally we search for a value of *g* that produces a fit to the observed value of *q* for the simulated random walk within the observational error.

An important point to note is that the three steps are done sequentially, and that the two parameters *f* and *g* are independent and uncorrelated. The model is therefore characterized by only the 2 uncorrelated free parameters that are determined from the observed scaling exponents.

As a practical matter, it typically takes only a few minutes of wall clock time to find acceptable models using a Monte Carlo or grid search algorithm that simultaneously fits the *b*-value, the *p*-value and the *q*-value of the observed California catalog within the observational error.

**Examples of ERAS Catalog Simulations and Comparison with California Data.** Figures 3-4, along with Figures A2-A3 in the appendix, illustrate some example details of ERAS simulated catalogs, and compare them to the observed catalog. In Figure 3, we show maps of seismicity for the period 1960-present in space and time for the observed California catalog (left) with an ERAS simulation (right). Seismicity is for a box of size 10º x 10º latitude-longitude centered on Los Angeles. Two large earthquakes are indicated, which are further examined in Figure 4. The earthquakes in the left panels are for the 1992 M6.2 Joshua Tree - M7.3 Landers earthquakes, the right panels are for a simulated ERAS M7.1 earthquake. Figure 4 shows the magnitude-time plot in the top panels, and the daily number of small earthquakes following those events in the bottom panels.

## Science Transformers: AI Enhanced Time Series Forecasting

In Fox et al. (2022), we reviewed previous approaches to nowcasting earthquakes and introduced new approaches based on deep learning using three distinct models based on recurrent neural networks and transformers. We discussed different choices for observables and measures presenting promising initial results for a region of Southern California from 1950–2020. Earthquake activity is predicted as a function of 0.1-degree spatial bins for time periods varying from two weeks to four years.

The overall quality is measured by the Nash-Sutcliffe efficiency criterion comparing the deviation of nowcast and observation with the variance over time in each spatial region. We found a reasonable degree of success nowcasting future seismicity, although anticipating the largest earthquake occurrence is still problematic. The software is available as open source together with the preprocessed data from the USGS

# QuakeGPT: A Generative Pretrained Earthquake Transformer

**Earthquake Transformers and QuakeGPT**. Our major focus is upon the development and deployment of the Earthquake Generative Pretrained Transformer model "QuakeGPT" as a means of predicting the future evolution of the nowcasting time series curves. The workflow summary is shown in Figure 5.

As explained in the seminal paper by Vaswani et al. (2017), a transformer is a type of deep learning model that learns the *context* of a set of time series by means of tracking the relationships in a *sequence of data,* such as the words in a sentence. Pretrained transformers are the foundational technology that underpins the new AI models ChatGPT (Generative Pretrained Transformers) from openAI.com, and Bard, from Google.com. In our case, we hypothesize that a transformer might be able to learn the *sequence of events* leading up to a major earthquake. A schematic illustration of the attention-based science transformer model is shown in Figure 6.

Typically, the data used to train the model is in the billions or larger, so these models, when applied to earthquake problems, need the size of data sets that only long numerical earthquake simulations can provide. In this research, we begin to develop the Earthquake Generative Pretrained Transformer model, QuakeGPT, in a similar vein.

Among the tasks we are currently pursuing is the definition of optimal methods to define feature vectors for the data, together with optimal approaches to combine the statistical seismicity simulations with the dynamical earthquake simulations, thus defining the training data that will be used for the QuakeGPT models. We are also investigating the length and complexity that simulation data sets need to be, given the comparatively much smaller amount of observational data that is available.

We now discuss the components of the QuakeGPT model. These include:
- A Generative ERAS models for training the transformer model;
- Defining the architecture of the transformer model;
- Building long simulations for training the transformer;
- Testing and validating the QuakeGPT model on simulation data;
- Optimizing the G-ERAS model and applying the result to the observed catalog data in northern California.

**ERAS Earthquake Simulations.** As we have discussed, generative ERAS models are easy to code and can be used to generate catalogs of arbitrary length. They are well suited to applications involving machine learning, and by systematically varying their properties and parameters, they can be used to formulate a variety of null hypotheses and alternate models. As an example, we showed a representative ERAS model for a box of size $10^o$ x $10^o$ centered on Los Angeles in Figure 3. These simple models can also be used as "testbeds" for a variety of machine learning applications, since they can be constructed to contain most, if not all, of the statistical properties of observed catalogs.

**Transformer Architecture.** The transformer contains the following components:
- *Encoder:* The Transformer model uses multiple encoder layers, each consisting of self-attention and feed-forward neural networks. These layers process the input sequence and capture the temporal dependencies between the data points (Figure 6).
- *Positional Encoding*: Before passing the data through the encoder layers, positional encoding is added to the input sequence. It provides information about the order of the data points, as the Transformer model does not have an inherent notion of sequence order. However, we note that in the time series prediction problem, temporal ordering is implicit in the input data.
- *Global Average Pooling*: After the encoder layers, global average pooling is applied to aggregate the information from all encoded data points into a fixed-size representation.

- *Decoder Linear Layer (Output):* A linear layer is applied to the pooled representation to map it to the output dimension.
- *Predictions:* The output of the linear layer represents the predictions made by the Transformer model for the next data point in the time series. For times beyond the end of the validation data set, these predicted data points are fed back into the model as inputs to predict yet more future values of the data. It is clear that this process will accumulate errors rapidly as uncertain data values are used to predict future data values. Therefore, predictions too far into the future are probably not useful.

Initially, we use the following parameters, following which we will explore the effect of variations in these parameters. We use code from PyTorch, with the Adam Optimizer and MSE Loss function:

- Input Dimension: 36 (months)
- Output Dimension: 1 (month)
- Hidden Dimension: 32 (Neurons/Layer)
- Number of Encoder/Decoder Layers: 2
- Number of Self-Attention Heads: 4
- Dropout: 0.2
- Learning Rate: 0.001
- Model Size: 170KB

**Queries, Keys and Values**. In general, attention-based transformers use the concepts of queries, keys and values. In our case of time series prediction, we use feature vectors consisting of 36 months of data to predict the next value, thus a one-step walk ahead predictor. We identify the "keys" with the 36-months of data in the training data set, and the corresponding "values" as the value of the next data point. We identify the "queries" as the 36-month feature vectors in the validation data set that are used to predict the subsequent value of the time series. By analogy, the "context vector" is the predicted future value of the time series in the validation data.

While nowcasting provides us with a method for computing probabilities of major earthquakes currently and in the near future, there remains a critical need to accurately extrapolate the nowcast curves into the future, so that large earthquakes can be anticipated farther into the future. The nowcasting methodology, combined with science transformers, provides us with the mechanism to accomplish this objective, albeit with accumulating errors.

As an example, our initial results shown here in Figure 7 are for the California box centered on Los Angeles, shown in Figure 3. These results are based on training the transformer model on 2,021 years of ERA5 data, then applying it to 53 years of independent test ERA5 data (cyan region), a time span similar to the observed data in California as shown in Figures 1-2. Then predictions were made for data values *beyond the test data* (magenta region, Figure 7) by feeding the previously predicted output values into the transformer as inputs to predict future values as described previously.

Recall that the nowcast time series represents an exponentially smoothed average of the monthly rate of small earthquakes. In addition, the number of small earthquake aftershocks can be used to estimate the implied magnitude of the mainshock that might have produced this number of aftershocks. Therefore, once the algorithm predicts the value of the nowcast time series beyond the known data, the last step would be to invert the time series to calculate the implied number of small earthquakes, and from that, the implied magnitude of the mainshock that produced this number of small earthquakes. However, we defer this calculation to a future publication.

We are particularly interested in events that are associated with a sudden large change in the time series, a sudden decrease in the earthquake potential state, associated with a large sudden activation of small earthquakes. Performing this inversion is in fact straightforward and will be an eventual part of the QuakeGPT model whose goal will be to calculate the implied magnitudes of these large mainshocks.

## Summary and Discussion

In this paper, we have discussed a new approach to earthquake nowcasting that relies on the construction and use of an Earthquake Generative Pretrained Transformer model, QuakeGPT. The generative pretrained transformer formalism employs a similar approach to that which is used in the ChatGPT model available from Openai.com, with the difference being that the training data sets are simulation catalogs of earthquake data, rather than a very large corpus of text on the internet.

We began with a summary of our current nowcasting methods, and then introduced a new stochastic simulation model "ERAS" (Earthquake Rescaled Aftershock Simulation) that we used to train the transformer. Next we discussed the construction of a dot product attention-based science transformer "QuakeGPT" that we use in the AI-enhanced nowcasting model. We also described the queries, keys and values that are used in the transformer technology.

We then showed that a model can be trained using several thousand years of simulation data, and showed how the model can fit an independent validation time series, predicting one-step walk-ahead values for the time series. Next, using the predicted values as input, we show that it is possible to predict future values of the time series beyond the end of the validation set, albeit with increasing levels of prediction error. Finally we assess the overall prospects and future promise for use of QuakeGPT as a means of anticipating the occurrence of future large earthquakes.

As was implied in Figure 5, we can also in principle use physics-based simulations of interacting faults to train QuakeGPT. It should be pointed out, however, that using these fault-based models entails a higher level of difficulty in ensuring that the training model accurately represents the fault system dynamics (e.g., Van Aalsburg et al, 2007). Models of this type such as Virtual Quake (Rundle et al., 2004) or RSQSim (Richards-Dinger and Dieterich, 2012) are based on thousands of parameters, are expensive to run, and make acquiring the necessary large volume of training data correspondingly difficult.

Given the rapid recent development of transformer technology based on self-attention, it can be anticipated that models of the type we discuss here will be increasingly employed for earthquake nowcasting. We should also note that further developments will include spatially localizing the time series so that spatial nowcasts in small areas can be carried out, thus improving the information imparted by the nowcast. We plan to continue the development of these projects and will report on those developments in due course. We have not applied the transformer technology to the observed earthquake data set for reasons of caution in disseminating preliminary results, but we plan to address this problem in the near future.

**Acknowledgements**. Research by JBR and IB was supported in part under DoE grant DE-SC0017324 to the University of California, Davis. JBR would also like to acknowledge generous support from a gift to UC Davis by John Labrecque. The authors would also like to acknowledge a helpful review by an anonymous referee.

**Open Research.** Python code that can be used to reproduce the simulation results of this paper can be found at Rundle (2024b). Data for this paper was downloaded from the USGS earthquake ComCat catalog for California, and are available there. Python code at Rundle (2022b) can be used to download and model these data for analysis using the methods of Rundle et al. (2022a). Please see https://doi.org/10.5281/zenodo.11630114 for ERAS code. For the QuakeGPT transformer code, visit: https://doi.org/10.5281/zenodo.11630546

**Data.**  Data for this paper was downloaded from the USGS earthquake catalog for California, and are freely available there. The Python code mentioned above can be used to download these data for analysis.

# Appendix: Building the ERAS Model

**Initial Model Set-up and Event Classification.**  Given a USGS catalog with completeness magnitude $\mu$, we create a catalog with the same exact number of events, having completeness magnitude $\mu$, and with events drawn from a Gutenberg-Richter magnitude-frequency distribution with the observed *b*-value. Initially, the time between these events $\Delta t$ is set using a Poisson (exponential distribution) for inter-event times $\Delta t$:

$$\Delta t = -Log(1 - \zeta) \tag{A1}$$

where $\zeta$ is a random deviate drawn from a uniform distribution. Once the catalog has been defined, the times are proportionately rescaled (expanded or compressed) to lie between an observed catalog start date (e.g., such as 1/1/1990) and the catalog final date (e.g, such as the present 3/1/2024).

Next, the Bath's number $\Delta m_B$, which is the typical difference between the mainshock magnitude and the largest aftershock magnitude, is used to estimate the number of aftershocks from a mainshock. Given a mainshock magnitude *m*, we label the next $N_a$ events in time as "aftershocks" ("a"), where $N_a$ is given by:

$$N_a = 10^{(m - \mu - \Delta m_B)} \tag{A2}$$

Events not labeled as aftershocks are labeled as "background" events ("b"). Background events having aftershocks are considered to be "mainshocks".

An implication of the above classification is that only background events with magnitudes $m \geq \mu + \Delta m_B$ can have aftershocks.

**Rescaling Time Intervals for Aftershock Power Law Scaling: Parameter *f*.**  We note that the existence of the Omori law implies, like other fractal power laws, a geometric scale invariance. The next step is to identify cycles of "mainshock" events in order of descending magnitude. Finally, a geometric rescaling of the time intervals between the "aftershocks" and the "mainshocks" is applied in each cycle using a scale-invariant algorithm. The process is illustrated in Figure A1 and described in the following steps.

**Rescaling Spatial Migration for Aftershock Power Law Scaling: Parameter *g*.**

A. **Initial Model Set-up and Event Classification:**

- Given a USGS catalog with completeness magnitude $\mu$. We create a catalog with the exact same number of events, having completeness magnitude $\mu$, and with events drawn from a Gutenberg-Richter magnitude-frequency distribution with the observe *b*-value.

- Initially, the time between these events $\Delta t$ is set using a Poisson (exponential distribution) for inter-event times. Once the catalog has been defined, the times are proportionately rescaled to lie between an assumed start date (e.g., such as 1/1/1970) and an assumed final date (e.g, such as the present 2/12/2024).

- The Bath's number $\Delta m_B$, which is the typical difference between the mainshock magnitude and the largest aftershock magnitude, is used to estimate the number of aftershocks from a mainshock. Given a mainshock magnitude $m$, we label the next $N$ events in time as "aftershocks" ('a'), where $N$ is given by equation (A2).

- Events not labeled as aftershocks are labeled as "background" events ('b'). Background events having aftershocks are considered to be "mainshocks".

- An implication of the above classification is that *only* background events with magnitudes $m \geq \mu + \Delta m_B$ can have aftershocks.

### B. Rescaling Time Intervals for Scale-Invariance: Parameter $f$

- We begin by noting that the existence of the Omori power law implies, like other fractal power laws, that at the heart of the power law is a geometric scale invariance.

- We therefore identify cycles of events in order of descending magnitude, then we carry out a geometric rescaling of the time intervals between the "aftershocks" and the "mainshocks" in each cycle in a proportional, scale-invariant way. The process is illustrated in Figure 1.

- For example, assume that the largest magnitude event in the simulation is $m_L$. Then the largest magnitude aftershock is expected to have magnitude $m_a = m_L - \Delta m_B$. We identify all events having $m \geq m_a$, and consider the intervals between these events as a set of $m_m$-"mainshock" cycles.

- We then proportionately rescale all the earthquakes between these $m_m$ events as in Figure 1.

- The rescaling factor $f < 1$ is shown in Figure 1, and is applied to the n events between successive $m_m$ events.

- For example, if there are 10 events, indexed as 0,...,9 in the interval between the two magnitude $m \geq m_m$ events, the rescaling factor for the event immediately following the first $m \geq m_m$ event (the first "aftershock") is $f^9$. For the last "aftershock" event, $f^0 = f$.

- Correspondingly, if the original, unclustered time intervals between the $m_m$ mainshocks are indexed as $\Delta t_9$ and $\Delta t_0$, for example (Figure A1), the adjusted time intervals are $\Delta t'_9 = f^9 \Delta t_9$ and $\Delta t'_0 = f^0 \Delta t_0$. Clearly then, $\Delta t'_9 < \Delta t_9$ and $\Delta t'_0 = \Delta t_0$.

- The adjusted time intervals are then proportionately smaller than the original time intervals as a result of this geometric rescaling algorithm. Their time ordering remains unchanged.

- The rescaling algorithm is then applied recursively. After rescaling at the $m \geq m_m$ level, we repeat this procedure at the level $m \geq m_m - 0.1$, in other words at a level 0.1 magnitude unit lower. This recursive procedure continues to be applied down to the completeness magnitude $\mu$.

### C. Spatial Event Locations: Parameter $g$

As described above, we model the aftershock migration away from the mainshock by a random walk. As is well known, for a 2-dimensional Brownian random walk is characterized by the simple relation:

$$R^2 = N \epsilon^2 \tag{A3}$$

where $R$ is the average radius of the walker after $N$ steps, and $\epsilon$ is the step size.

We analyze the observed catalog to find $\epsilon_{Lat}$ and $\epsilon_{Long}$, and then determine the value of a parameter $g$ by fitting to the observed Omori-Utsu spatial scaling law with parameter $q$. We apply the random walk scaling relation by examining the aftershock locations for stacked mainshocks, and compute the distance migrated in both latitude and longitude as a function of the aftershock number.

We do this for each mainshock, and then average the results. In the algorithm, we accept an aftershock location with 95% confidence if the computed step size for either latitude or longitude is within 2 standard deviations of the expected values for latitude and longitude, based on the current number of steps in the walk.

Once we have found $\epsilon_{Lat}$ and $\epsilon_{Long}$ by analyzing the observed catalog, we proceed to generate a random walk around each mainshock with step sizes $\epsilon_{Lat}$ and $\epsilon_{Long}$, using the number of aftershocks $N$ for the number of steps in the random walk. We then vary the step sizes as $g\,\epsilon_{Lat}$ and $g\,\epsilon_{Long}$ until we find a value of $g$ leading to a $q$ value that fits the data within the observational error. The result for the ERAS catalog discussed in this paper is shown in Figure A3, and compared to the corresponding $q$-value data for the California catalog.

### D. Constructing the Catalog

As discussed above, to build the catalog we start by generating a catalog with random interevent times, and magnitudes drawn from a Gutenberg-Richter distribution. These catalogs have exactly the same number of events as the observed catalog. Catalogs are continually generated until a catalog with a $b$ value within the margin of error of the observed catalog is found. Next, the random times are compressed or expanded so that the catalog has the same total elapsed time as the observed catalog.

At this point, the geometric rescaling of aftershock times is carried out. Since the overall elapsed time will change, the event times are again compressed or expanded so that total elapsed time is the same as the observed. To select locations for the events, we allow all aftershock events to migrate outward from the mainshock epicenter using step sizes $g\,\epsilon_{Lat}$ and $g\,\epsilon_{Long}$. Epicenters for "background events" are selected randomly from the epicenters of earthquakes in the observed catalog. In addition, large events, for example having $m \geq 6$, are assigned epicenters where such large earthquakes actually occurred.

As the next step, we identify all small events having a magnitude less than the value of magnitude $m < \mu + \Delta m_B$ that are labeled as background ('b'), not aftershocks ('a'). These events are removed from the catalog and copied to a temporary file containing only background events. At this stage, there is now a "triggered event" file and a "background event" file.

In the final steps, the time in the background file are randomized by randomly assigning times from the file (selection without replacement) to lat-long-depth locations in the same file. Next the background event file with randomized times is appended to the triggered event file, and the resulting combined catalog is sorted so that times are in ascending order. The result of these computations is shown in Figures A2 and A3.

As note above, parameters $f$ and $g$ are found by carrying out a grid search for a catalog whose $p$-value and $q$-value data fit the observed values. The result for the ERAS catalog discussed in this paper is shown in Figure A3, and compared to the corresponding $p$-value and $q$-value data for the California catalog. For the $p$-values, we fit the Omori-Utsu temporal scaling data in the range of 1 to 20 days, and for the $q$-value data, we fit the Omori-Utsu spatial scaling data in the range of 4 to 30 KM.

# Figures

**Figure 1.** Illustration of the construction of the nowcast model, a 2-parameter filter on the small earthquake seismicity (see Rundle 2022, 2023 for details). a) Seismicity in the Los Angeles region since 1960, M>3.29. b) Monthly rate of small earthquakes as cyan vertical bars. Blue curve is the 36 month exponential moving average (EMA). c) Mean rate of small earthquakes since 1970. d) Nowcast curve that is the result of applying the optimized EMA and corrections for time varying small earthquake rate to the small earthquake seismicity. e) Optimized Receiver Operating Characteristic (ROC) curve (red line) used in the machine learning algorithm. Skill is the area under the ROC curve and is used in the optimization. Skill tradeoff diagram shows the range of models used in the optimization.

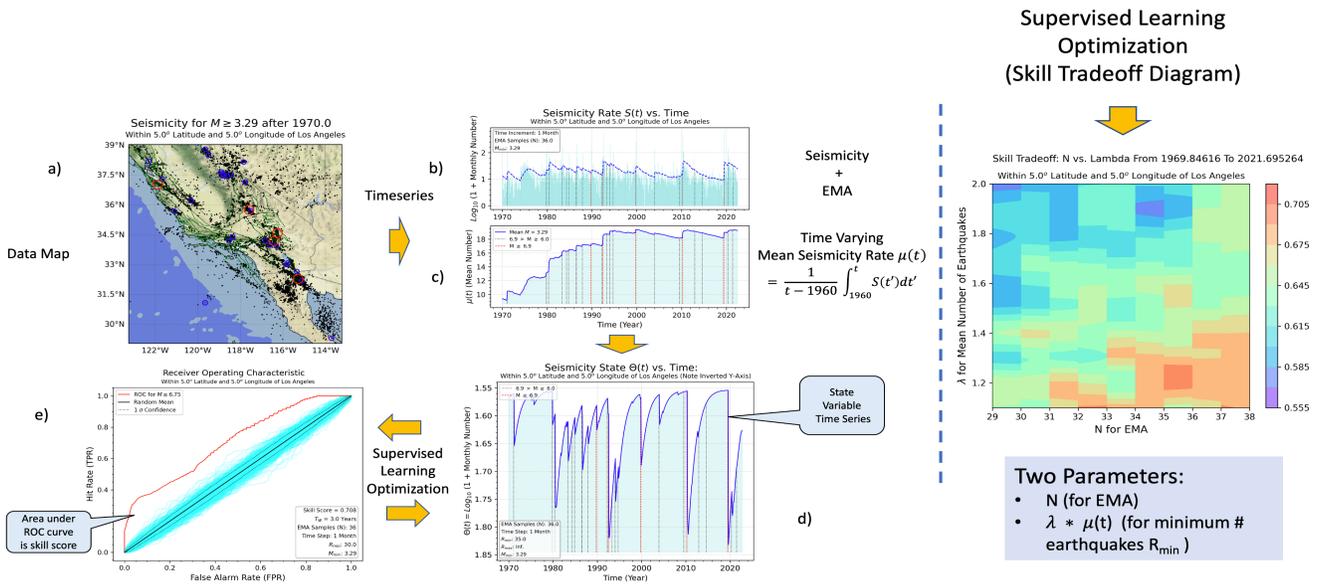

**Figure 2.** Frame from a movie of the nowcast for California earthquakes since 1970 (Rundle et al., 2022a; Rundle Movie 2024). Details of the various panels are shown in the text balloons in the figure and are discussed in the text.

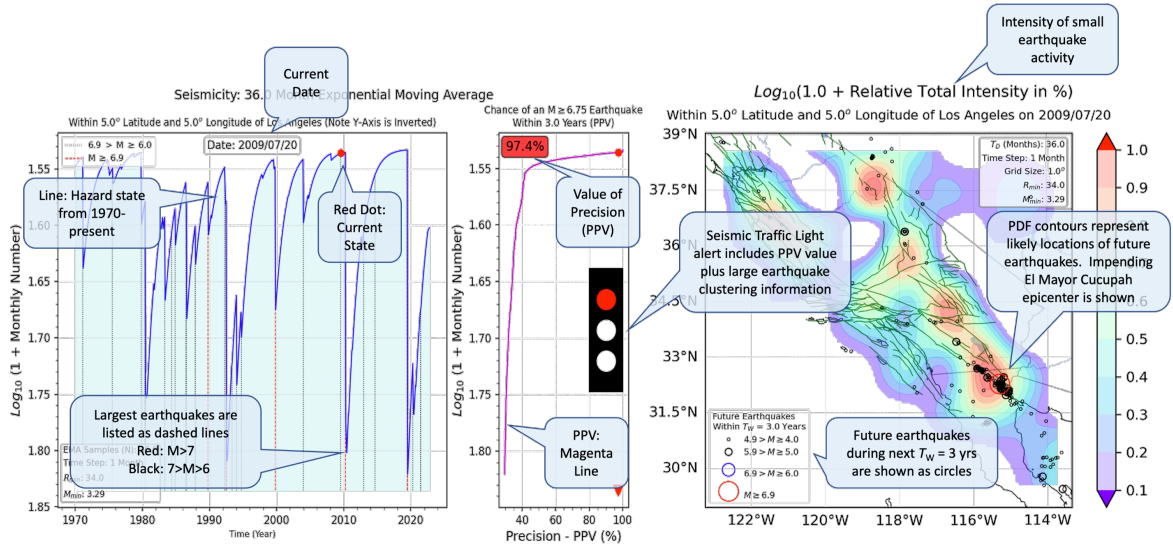

**Figure 3.** Comparison of maps of seismicity for the period 1960-present in space and time for the observed California catalog with the ERAS simulation. Seismicity is for a box of size 10º x 10º latitude-longitude centered on Los Angeles. The red circles denote the epicenters of large earthquakes M≥6.9 as shown in the figure insets. Two large earthquakes are indicated. Left: The M7.3 June 28, 1992 Landers earthquake in the

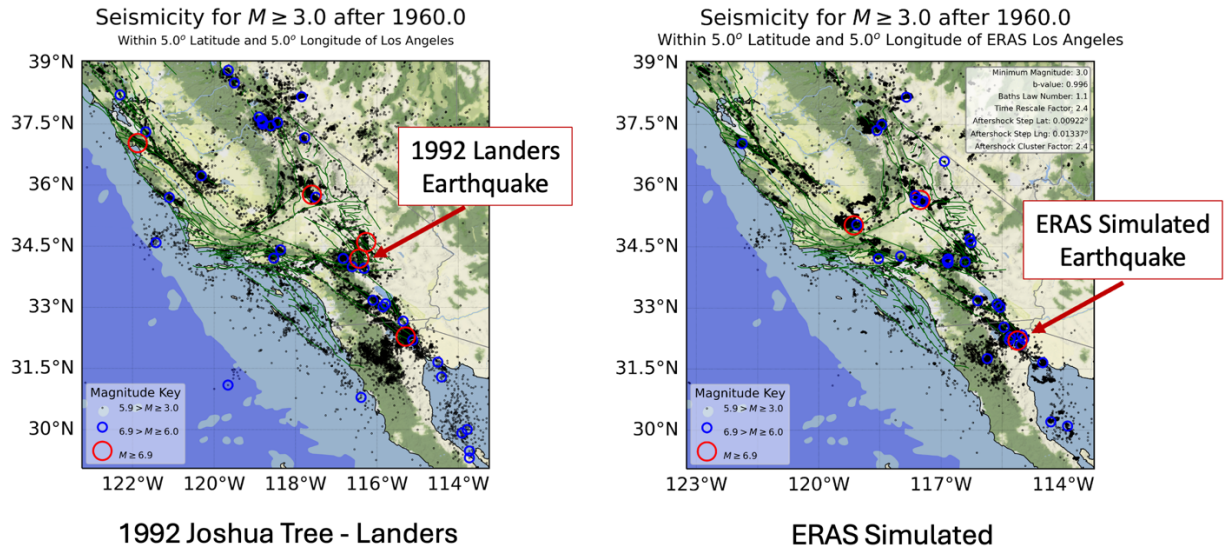

Mohave Desert. Right: An M7.1 ERAS simulated earthquake.

**Figure 4.** Comparison of details of the two large earthquakes shown in Figure 3. Left: Data for the M7.3 June 28, 1992 Landers earthquake and the M6.2 Joshua Tree earthquake on April 22, 1992. Right: An ERAS simulated mainshock with magnitude M7.1 on September 7, 2014, together with its aftershocks. Top: Magnitude vs Time for these events for a time period of 200 days. Bottom: Daily number of earthquakes M3.0 for a time period of 200 days. For the observed catalog, day 0 is April 22, 1992, the date of the Joshua Tree earthquake. For the simulated earthquakes, day 0 is the date of the mainshock, September 7, 2014.

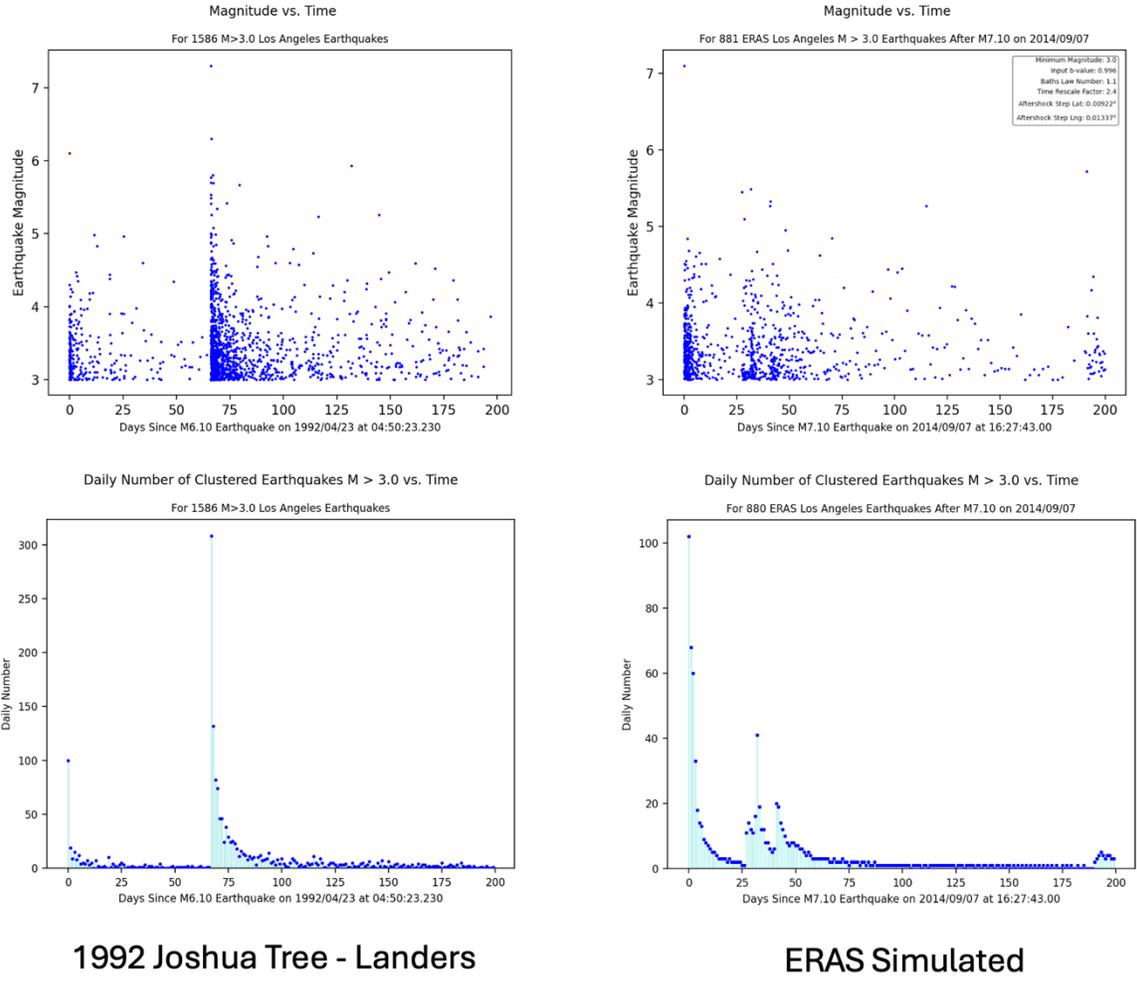

**Figure 5.** Workflow for the QuakeGPT model. Data is fed into the model in the form of physics-informed stochastic earthquake simulations (ERAS model), to add to physics-based dynamical fault models (not considered in this current paper). These long-time simulations are then used to pre-train the attention-based science transformer model. The pretrained transformer model is then used to predict the validation data, followed by prediction of values of the unobserved future time series. Potential hazard alerting can be carried out using a traffic-light or other alerting system.

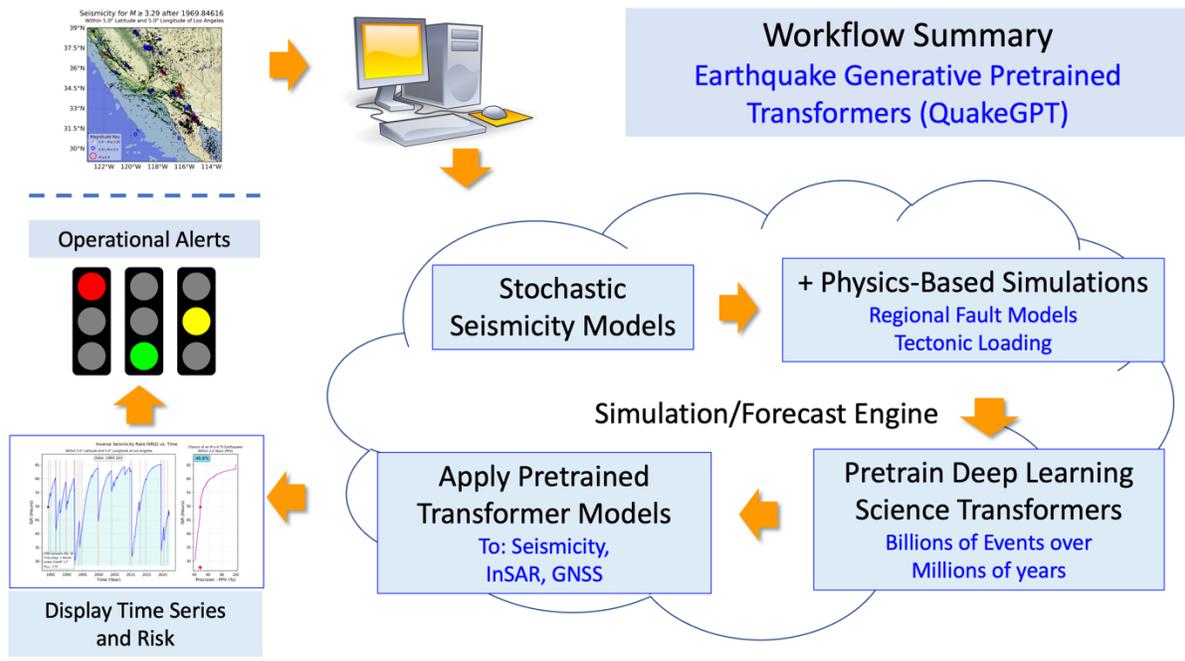

**Figure 6.** General structure of the attention-based science transformer code. Values of a time series are fed into an encoder, which consists of N layers, each with M self-attention heads. Positional encoding is then carried out (if needed), followed by one or more feed-forward layers. This is followed by a global average pooling layer, followed by a decoder layer that predicts the next value(s) of the time series.

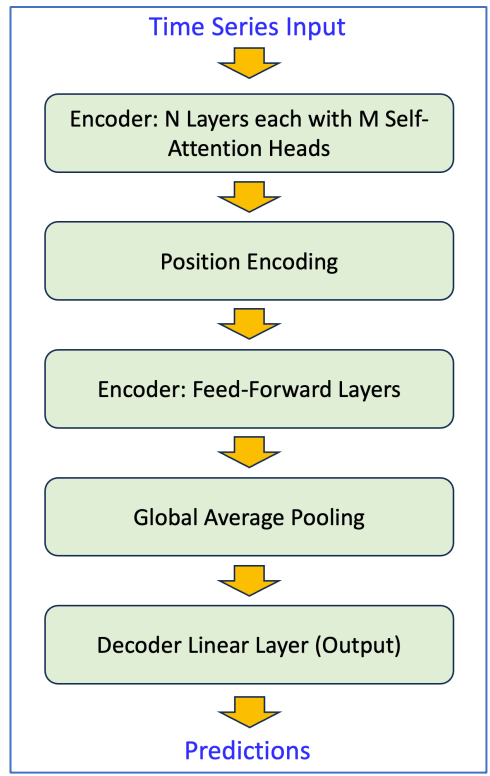

**Figure 7.**   Image showing the application of the trained transformer to an independent, scaled nowcast validation curve (green shading), followed by prediction of future values beyond the end of the nowcast curve (magenta shading). In this model, we use 6006 training epochs. 36 previous values are use to predict the next value. Dots show the predictions and the solid line shows the nowcast curve whose values are to be predicted. Green dots show the predictions of the transformer up to the last 37 values. 36 blue dots are predictions that were made and then fed back into the transformer to predict the final point (red dot). In this model, 50 members of an ensemble of runs were used to make the predictions. The dots represent the mean predictions. Brown areas represent the $1\sigma$ standard deviations to the mean values. In this model 2021 years of simulation data were used to train the model. The input dimension is 36 points, output dimension is 1 point (36 previous values are used to predict the next value). Hidden dimension of the neural network in the encoder-decoder layers is 32 neurons. Number of layers is 2 for this simple model, 1 encoder, 1 decoder. Number of self-attention heads is 4. Dropout rate is 0.02, learning rate for the gradient descent Adam optimizer is 0.001.

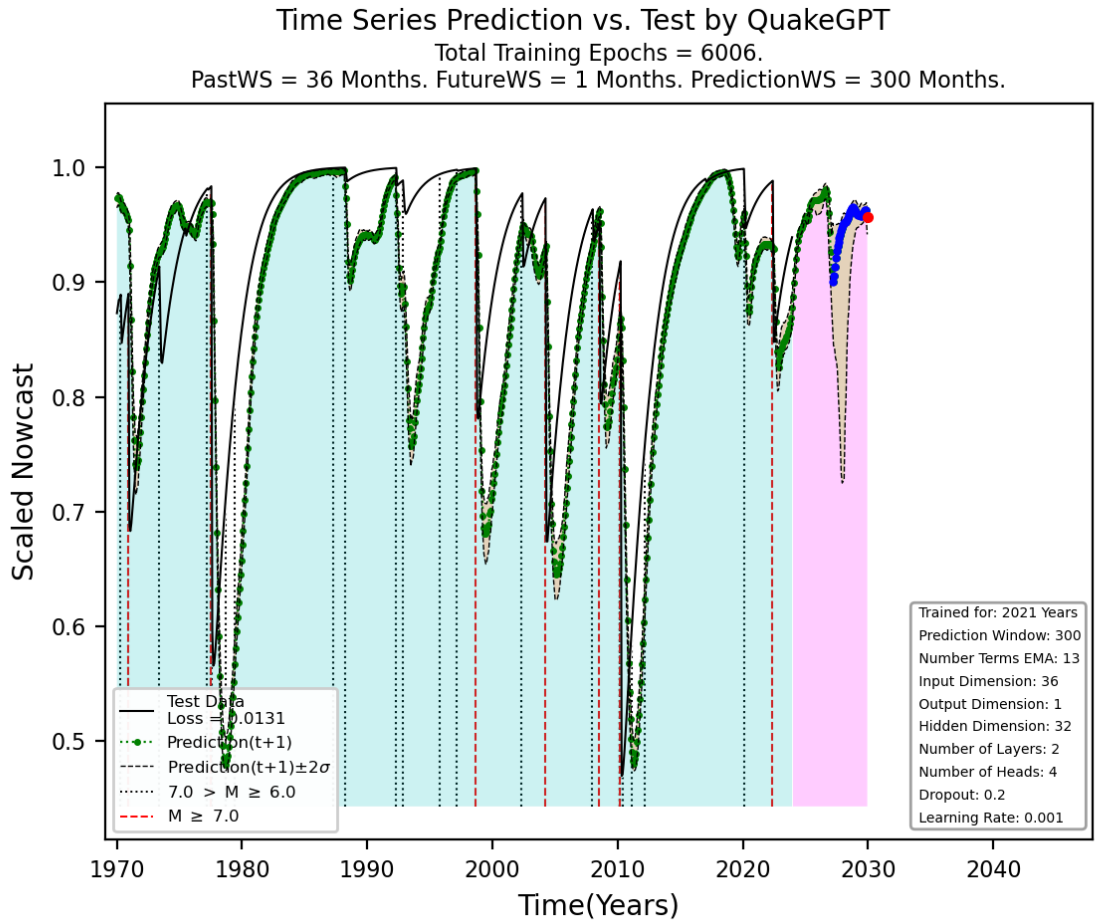

**Figure A1.** Schematic illustration of the re-clustering method described in the text. Given two earthquake "mainshocks" of some magnitude $m_{main}$, the smaller magnitude events between those mainshocks are identified. Using Bath's law, the subset of those smaller events that represent "aftershocks" are identified. Of the 3 smaller events shown, the first 2 were labeled as "aftershocks" ("a"), and the last event is labeled as "background" ("b"). These 2 "aftershocks" then have their times of occurrence altered to be earlier, and closer to the first mainshock. The time of the "background" event is unchanged. The "aftershocks" are moved closer to the earlier mainshock in time, using a geometric rescaling factor $f$ as shown in the figure.

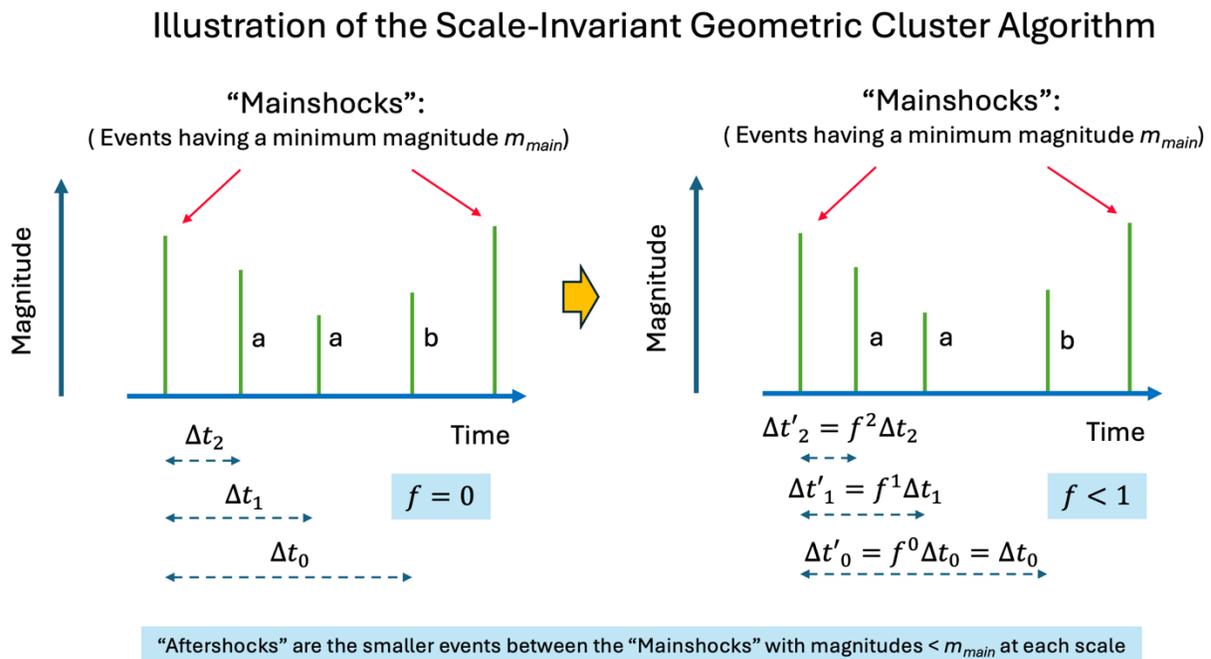

**Figure A2.** Comparison of observed California catalog with ERAS simulated catalog for the period 1960-present. Seismicity is for a box of size 10° x 10° latitude-longitude centered on Los Angeles. Left figures: Observed catalog data. Right figures: ERAS simulated data. Top figures: Magnitude vs. Time for the years 1960-present for magnitudes M>3. Bottom figures: Gutenberg-Richter magnitude-frequency distributions.

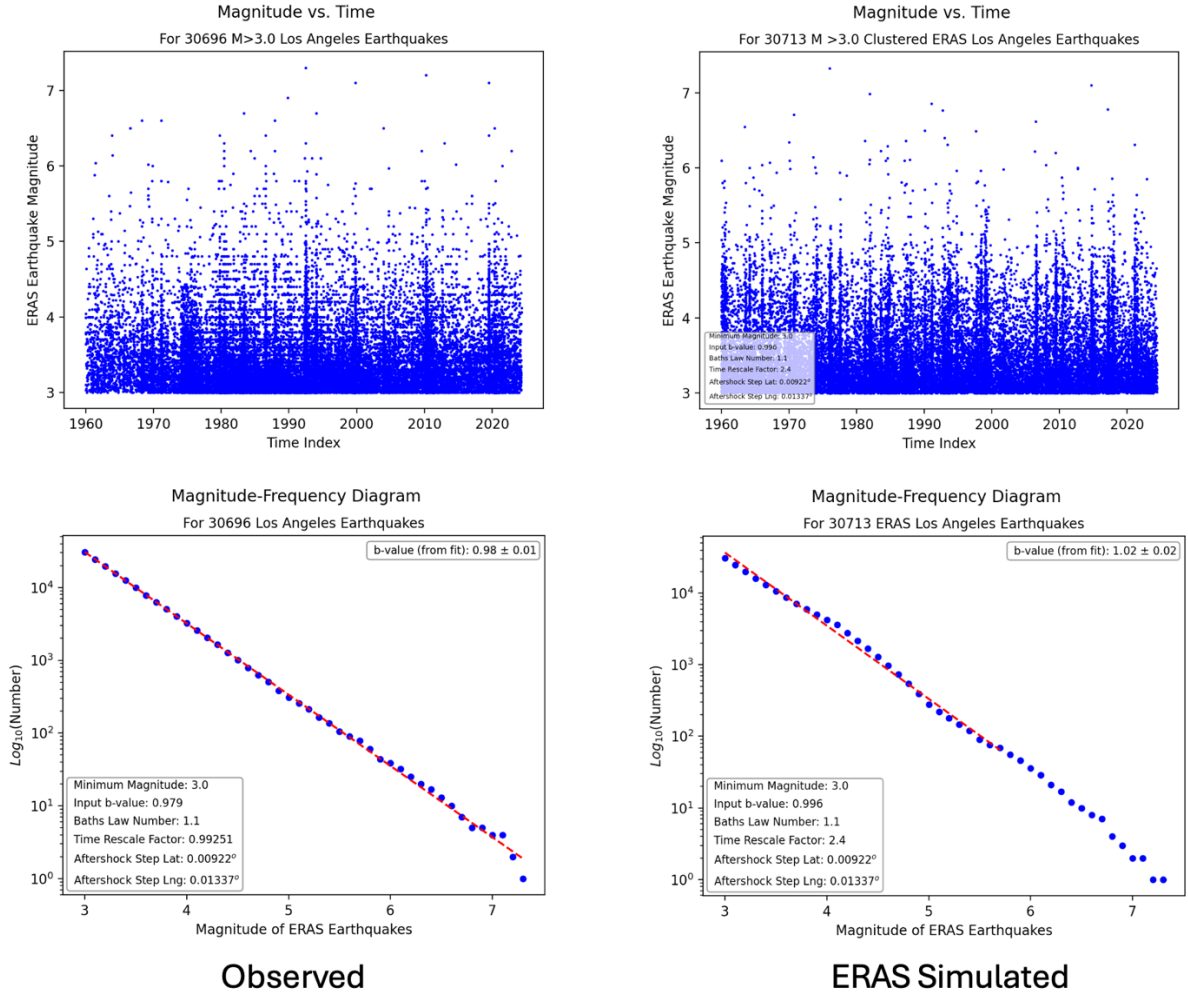

**Figure A3.** Comparison of statistics of aftershock decay in space and time for the period 1960-present for the observed California catalog with the ERAS simulation. Seismicity is for a box of size 10° x 10° latitude-longitude centered on Los Angeles. Left figures: Observed catalog data. Right figures: ERAS simulated data. Top figures: Omori-Utsu decay in time for stacked aftershocks for magnitudes M≥5.9. Bottom figures: Omori-Utsu decay in distance from the mainshock epicenter for stacked aftershocks for magnitudes M≥5.9.

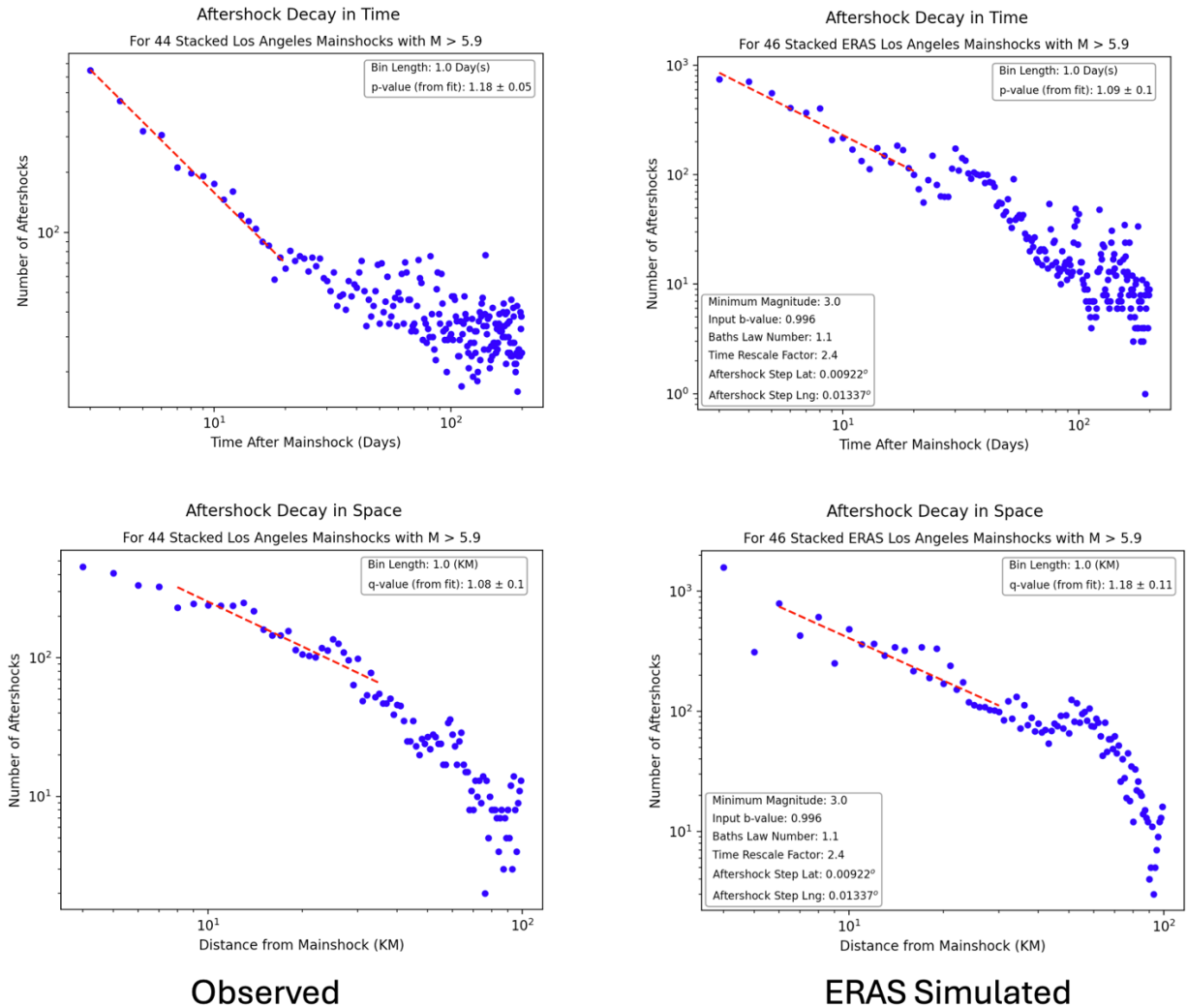